\documentclass[aps, preprint,prl, superscriptaddress]{revtex4-1}
\usepackage{titlesec}%
\usepackage{subfigure}%
\usepackage{amsmath}%
\usepackage{amsfonts}%
\usepackage{amssymb}%
\usepackage{feynmp}%
\usepackage{graphicx}%
\usepackage{setspace}%
\usepackage{comment}%
\usepackage{dsfont}%
\usepackage{color}%
\usepackage{bbold}
\usepackage{tikz}
\usetikzlibrary{arrows,calc}
\usepackage{hyperref}
\hypersetup{colorlinks,allcolors=black}

\newcommand{\ket}[1]{\left| #1 \right>} 

\usepackage{soul}   

\usepackage{xspace}
\usepackage{siunitx}
\usepackage{lineno}

\titleformat{\paragraph}
{\normalfont\normalsize\bfseries}{\theparagraph}{1em}{}
\titlespacing*{\paragraph}
{0pt}{3.25ex plus 1ex minus .2ex}{1.5ex plus .2ex}

\begin{document}

\title{Quantum Pair Generation in Nonlinear Metasurfaces with Mixed and Pure Photon Polarizations}

\author{Jiho Noh}
\affiliation{Sandia National Laboratories, Albuquerque, NM 87185, USA}
\affiliation{Center for Integrated Nanotechnologies, Sandia National Laboratories, Albuquerque, New Mexico 87185, USA}

\author{Tomás Santiago-Cruz}%
\affiliation{Max Planck Institute for the Science of Light, 91058 Erlangen, Germany}%
\affiliation{Friedrich-Alexander-Universität Erlangen-Nürnberg, 91058 Erlangen, Germany}%

\author{Vitaliy Sultanov}%
\affiliation{Max Planck Institute for the Science of Light, 91058 Erlangen, Germany}%
\affiliation{Friedrich-Alexander-Universität Erlangen-Nürnberg, 91058 Erlangen, Germany}%

\author{Chloe F. Doiron}
\affiliation{Sandia National Laboratories, Albuquerque, NM 87185, USA}
\affiliation{Center for Integrated Nanotechnologies, Sandia National Laboratories, Albuquerque, New Mexico 87185, USA}

\author{Sylvain D. Gennaro}
\affiliation{Sandia National Laboratories, Albuquerque, NM 87185, USA}
\affiliation{Center for Integrated Nanotechnologies, Sandia National Laboratories, Albuquerque, New Mexico 87185, USA}

\author{Maria V. Chekhova}%
\affiliation{Max Planck Institute for the Science of Light, 91058 Erlangen, Germany}%
\affiliation{Friedrich-Alexander-Universität Erlangen-Nürnberg, 91058 Erlangen, Germany}%

\author{Igal Brener}
\affiliation{Sandia National Laboratories, Albuquerque, NM 87185, USA}
\affiliation{Center for Integrated Nanotechnologies, Sandia National Laboratories, Albuquerque, New Mexico 87185, USA}
\email{ibrener@sandia.gov}

\date{\today}

\maketitle

\textbf{Abstract}

Metasurfaces are highly effective at manipulating classical light in the linear regime;  however, effectively controlling the polarization of non-classical light generated from nonlinear resonant metasurfaces remains a challenge. Here, we present a solution by achieving polarization engineering of frequency-nondegenerate biphotons emitted via spontaneous parametric down-conversion (SPDC) in GaAs metasurfaces, where quasi-bound states in the continuum (qBIC) resonances were utilized for boosting the biphoton generation.
By performing a comprehensive polarization tomography, we demonstrate that the polarization of the emitted photons directly reflects the qBIC mode's far-field properties.
Furthermore, we show that both the type of qBIC mode and the symmetry of the meta-atoms can be tailored to control each single-photon polarization state, and that the subsequent two-photon polarization states are nearly separable, offering potential applications in the heralded generation of single photons with adjustable polarization.
This work provides a significant step towards utilizing metasurfaces to not only generate quantum light but also engineer their polarization, a critical aspect for future quantum technologies.

\newpage

\begin{figure}[t!]
\centering
\includegraphics[width=12cm]{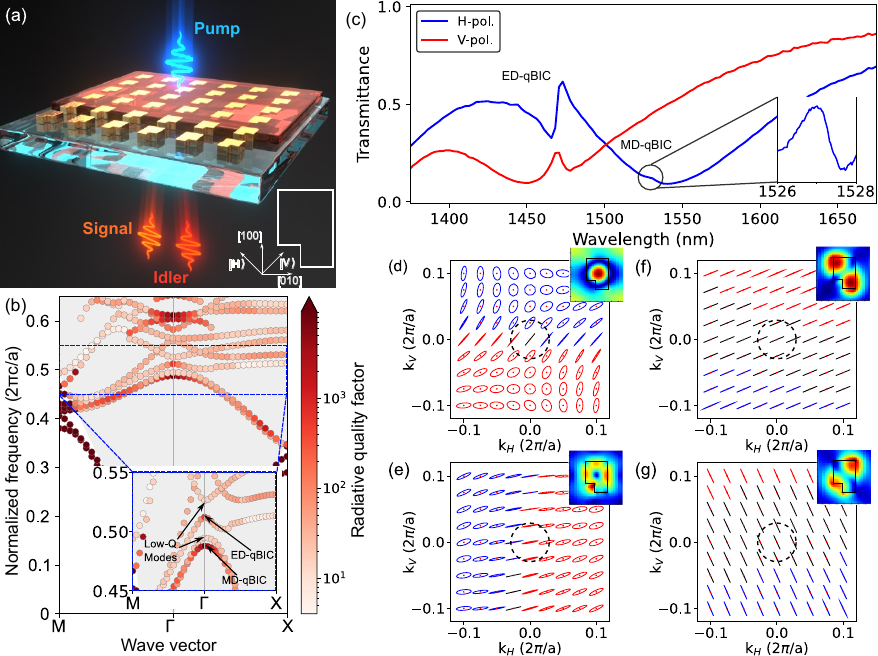}
\caption{(a) Schematic diagram of photon-pair generation in a quantum optical metasurface. The inset shows the orientation of the meta-atoms with respect to the crystallographic axes. (b) Band structure with the Q-factor along high-symmetry lines M-$\Gamma$-X calculated using guided-mode expansion method (GME) simulation~\cite{Minkov_ACSPhoton_2020}. Here, $a$ is the period of the lattice. (c) Measured white light transmittance for the incident polarization along the H-polarization axis (blue) and V-polarization axis (red) (d) The far-field polarization states of the metasurface calculated using GME simulation for the bands on which ED-qBIC and (e) MD-qBIC and (f-g) low-$Q$ modes lie at the $\Gamma$ point, respectively. The blue (red) ellipses represent right-handed (left-handed) polarization states. Dashed circles indicate the range in $k$-space covered by the effective collection NA of $\sim 0.03$. The insets in (d-g) show cross-sectional electric fields calculated using Lumerical at the center of the meta-atom for ED-qBIC, MD-qBIC and two low-$Q$ in-plane MD modes, respectively.} 
\label{fig:Schematic}
\end{figure}

The capability to generate and manipulate entangled photon pairs stands as a fundamental cornerstone for the progress and advancement of optical quantum technologies.
Traditional methods of quantum state engineering, which rely on spontaneous parametric down-conversion (SPDC) and spontaneous four-wave mixing, face severe restrictions owing to the simultaneous requirements for energy and momentum conservation (i.e., the phase-matching condition). 
SPDC is a quantum nonlinear optical process in which a higher-frequency pump photon, through a nonlinear interaction with a second-order nonlinear material, spontaneously splits into a pair of lower-frequency photons (a biphoton), often termed signal and idler photons [Fig.~\ref{fig:Schematic}(a)].
One promising approach to overcome the momentum constraint in SPDC is the use of ultrathin materials~\cite{Okoth_PRL_2019}.
This approach paves the way for new interesting concepts and applications, such as the broadband emission of photon pairs featuring subcycle correlation times and subwavelength correlation distances~\cite{Santiago-Cruz_OptLett_2021} and a giant degree of continuous-variable entanglement~\cite{Okoth_PRL_2019}, the generation of photon pairs with tunable polarization entanglement~\cite{Sultanov_OptLett_2022, Weissflog_arxiv_2023}, photon pair emission in quantum optical metasurfaces (QOMs)~\cite{Santiago-Cruz_NanoLett_2021}, and more recently SPDC sources based on organic materials, such as electrically-tunable liquid crystals~\cite{Sultanov_Nature_2024}, and  plasmonic metasurfaces strongly coupled to an epsilon-near-zero material~\cite{Jia_arXiv_2024}.
In particular, subwavelength QOMs can offer unparalleled opportunities for engineering the two-photon quantum state owing to their precisely controlled optical response.
Since the pioneering observation of SPDC from QOMs in 2021~\cite{Santiago-Cruz_NanoLett_2021}, remarkable progress has been achieved in this direction.
Scientists have demonstrated the generation of frequency-multiplexed~\cite{Santiago-Cruz_Science_2022}, spatially-entangled~\cite{Zhang_SciAdv_2022}, and bidirectional~\cite{Son_Nanoscale_2023,Weissflog_Nanophotonics_2024} photon pairs via SPDC in QOMs, including high-dimensional entanglement with multiphoton-state generation~\cite{Li_Science_2020}. 
Notably, these advancements have been facilitated by the field enhancement and control of density of states within QOMs, leading to boosted conversion efficiencies.
While significant progress has been made in manipulating the spectral and spatial degrees of freedom of photon pairs generated by QOMs, control over their polarization degree of freedom remains relatively unexplored.
In particular, the polarization states generated by the qBIC resonances were never studied.
Meanwhile, one recent study by Ma et al. developed a QOM based on a SiO$_2$ grating atop a subwavelength-thick LiNbO$_3$ layer, achieving emission of photon pairs with polarization dictated by the grating orientation \cite{MA_NanoLett_2023}, rather than by the inherent nonlinear tensor of the LiNbO$_3$ crystal, as observed in ultrathin crystals~\cite{Sultanov_OptLett_2022}. In addition, W. Jia et al  demonstrated the generation of a Bell state from a plasmonic metasurface where the polarization engineering stems from the nonlinear tensor of the material \cite{Jia_arXiv_2024}.
However, further understanding of the polarization characteristics of spontaneously generated photon pairs from QOMs is essential for unlocking their full potential in quantum technologies as polarization serves as a fundamental degree of freedom for encoding quantum information and preparing the desired quantum states for various quantum technologies. The integration of SPDC with nonlinear metasurfaces not only enhances our capacity to manipulate quantum states but also holds the potential for developing efficient quantum light sources with tailored functionalities.

In this work, we perform complete polarization tomography of frequency-nondegenerate photon pairs emitted via SPDC in QOMs based on meta-atoms fabricated from a $(001)$-oriented GaAs wafer~\cite{Santiago-Cruz_Science_2022}.
We employ single- and two-photon polarization tomography to characterize the quantum state and the purity of the emitted photons.
We demonstrate that by shaping the polarization of the resonant mode driving the nonlinear interaction, we can manipulate the polarization state of the photon pairs.
Furthermore, we show that the high polarization selectivity of the high-$Q$ resonance enables us to achieve a pure single-photon polarization state.
Finally, we also demonstrate that the two-photon polarization state is nearly separable, a direct consequence of the well-defined single-photon polarization states dictated by the resonant mode polarization.

To demonstrate the polarization manipulation of photon pairs generated via SPDC in QOMs, from a $(001)$-oriented GaAs wafer we fabricated metasurfaces made of square arrays of broken-symmetry meta-atoms [Fig.~\ref{fig:Schematic}(a)], similar to those studied in our previous paper~\cite{Santiago-Cruz_Science_2022}.
Energy conservation ensures the frequency anti-correlation of single photons forming a pair, while the longitudinal momentum conservation is relaxed in these subwavelength-thick metasurfaces.
As confirmed by both guided-mode expansion method (GME) simulations [Fig.~\ref{fig:Schematic}(b)] and white light transmittance measurements [Fig.~\ref{fig:Schematic}(c)], these metasurfaces support a set of high-quality factor ($Q$) resonances -- specifically, quasi-bound states in the continuum (qBICs), where the first two modes correspond to out-of-plane electric dipole (ED) and magnetic dipole (MD) qBICs.
For these lattice and meta-atom dimensions, the metasurface can be regarded as a square lattice composed of square meta-atoms, which exhibit $C_{4v}$ symmetry, with a rectangular protrusion for the symmetry breaking. 
Since the qBICs originate from symmetry-protected BICs\cite{Hsu_NatMat_2016, Koshelev_PRL_2018}, the protrusion serves as a control mechanism, enabling manipulation of the $Q$-factors and polarization at the $\Gamma$-point.
In addition to these qBICs, the deformation from a perfect square meta-atom lifts the degeneracy between bright photonic modes (in-plane MD modes) that possess low $Q$-factors, as indicated in Fig.~\ref{fig:Schematic}(b), while maintaining modal and polarization orthogonality. These qBICs and low-$Q$ modes feature distinct far-field patterns, as shown in Figs.~\ref{fig:Schematic}(d-g).

\begin{figure}[t]
\centering
\includegraphics[width=12cm]{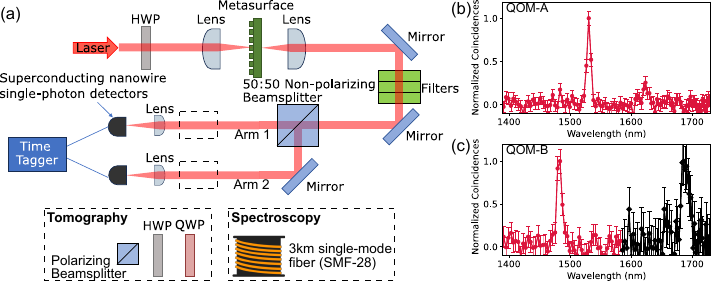}
\caption{(a) Experimental setup. A continuous-wave laser centered at 788.4 nm pumps the GaAs metasurfaces. The emitted SPDC radiation is filtered off the pump laser using a combination of longpass filters, and then sent to a Hanbury-Brown-Twiss-like interferometer for further analysis and detection. See main text for further information. (inset) Additional optical components and their positions in the setup for tomographic and spectroscopic measurements are indicated with dashed boxes. (b) SPDC spectra for MD-qBIC photons and their conjugate photons from QOM-A, measured in a single measurement. The difference in the heights of the peaks is due to the unequal detection efficiencies in the two arms. (c) SPDC spectra for ED-qBIC photons and their conjugate photons from QOM-B, measured in separate measurements, indicated in different colors (see SI).}
\label{fig:SPDC}
\end{figure}

We utilize a 15 mW, continuous-wave laser at 788.4 nm as a pump for SPDC, enabling the generation and measurement of frequency-nondegenerate photon pairs [Fig.~\ref{fig:SPDC}(a)].
The linearly polarized pump beam is controlled by a half-wave plate (HWP) before being focused (80 $\mu$m diameter) onto the metasurface using a lens with a 200 mm focal length.
The generated photon pairs are collected through a lens with a 100 mm focal length.
A series of long-pass filters (one at 1250 nm and two at 1400 nm) separates the collected photons from the pump beam.
Then, the filtered photons are directed to a Hanbury Brown-Twiss setup, where they pass through a 50:50 non-polarizing beam-splitter (NPBS) cube and are detected by superconducting nanowire single-photon detectors (SNSPDs).
A time tagger registers photodetection pulses from each SNSPD, and joint detection events are counted based on time differences to measure the rate of simultaneous photon detections, or coincidences. For detection, we couple the photons into single-mode fibers (SMF-28) using lenses with 18.4 mm focal lengths. The effective collection NA is $\sim 0.03$. 

We investigated two different QOMs to study the effect of each qBIC mode on the polarization state of the emitted photons.
Referred to as QOM-A [Fig.~\ref{fig:Schematic}(c)] and QOM-B [Fig.~\ref{fig:lin}(a)], these QOMs differ in meta-atom sizes and lattice constants. In this work, we focus on frequency nondegenerate SPDC driven by MD-qBIC and ED-qBIC, respectively.
Notably, qBIC resonances enhance the vacuum field at the signal wavelength and subsequently boost the rate of photon-pair generation, regardless of the pump wavelength.
Since both qBIC modes coexist, the QOMs can support two independent SPDC processes, each driven by its  qBIC\cite{Santiago-Cruz_Science_2022}.
To isolate the effect of a single qBIC, we judiciously selected the pump laser polarization and implemented additional spectral filters to guarantee no contribution from an unwanted qBIC mode (see Table \ref{tab:spectroscopy}), and measured the spectrum via fiber-assisted spectroscopy [see inset in Fig.~\ref{fig:SPDC}(a)].
In QOM-A, one photon of the pair is generated at the MD-qBIC resonance at 1527.7 nm, accompanied by its conjugate partner photon at the wavelength (1629.2 nm) determined by energy conservation [Fig.~\ref{fig:SPDC}(b)]. Similarly, in QOM-B [Fig.~\ref{fig:lin}(a)], one photon of the pair is generated around ED-qBIC resonance at 1484.0 nm with its conjugate partner photon generated at 1682.0 nm [Fig.~\ref{fig:SPDC}(c)].
For simplicity, we designate the photons generated at the qBIC resonances as ``signal photons" and their conjugate partners as ``idler photons".
The limited signal-to-noise ratio in the SPDC process due to background photoluminescence~\cite{Okoth_PRL_2019, Machulka_JPhysB_2014, Marino_Optica_2019} renders direct polarization measurements of single- or two-photon states impractical, necessitating full two-photon polarization tomography.

To characterize the polarization states of photon pairs generated via SPDC, we employed single-photon and two-photon polarization tomography~\cite{White_PRL_1999, James_PRA_2001, Sultanov_OptLett_2022, Sultanov_Nature_2024} and reconstructed the density matrices $\hat{\rho}$ representing the polarization states.
The state vector of an arbitrarily polarized single photon is a superposition of two basic states, $\left\{\ket{H},\, \ket{V}\right\}$:
\begin{equation}
    \ket{\Psi}=\alpha\ket{H} + \beta\ket{V},
\end{equation}
where complex coefficients $\alpha,\beta$  satisfy the normalization condition $|\alpha|^{2}+|\beta|^{2}=1$.
The single-photon polarization state cannot be determined by a single measurement.
To reconstruct the density matrix  describing the polarization state, we added different spectral filters to each arm to distinguish photons based on their frequencies and avoid further loss of information, which would reduce the measured purity of the states.
We then measured the coincidence rate under different polarization projections applied to the selected photon in the photon pair, simply by adding a quarter-wave plate (QWP), a HWP and a polarizing beamsplitter (PBS) cube in Arm 2 of the NPBS [see inset in Fig.~\ref{fig:SPDC}(a)].
This enables the projection of the selected photon onto a certain polarization state before its detection. 
Importantly, in this measurement no polarization filters were inserted in Arm 1; the corresponding detector was used only to `herald' the presence of the photon in Arm 2.
By measuring the coincidence rate with 6 different polarization projections in Arm 2 (Table \ref{tab:single_tomo}) and implementing the maximum likelihood state estimation algorithm~\cite{James_PRA_2001}, we can extract the density matrix $\hat{\rho}$ of the single-photon state. 
The reconstructed density matrix fully characterizes  the polarization state of single photons in Arm 2.
In addition to the density matrices, we can also extract the Stokes parameters and the corresponding degree of polarization from the aforementioned measurements as detailed in supplementary section II.

Furthermore, the two-photon state generated via nondegenerate SPDC is a ququart~\cite{Bogdanov_PRA_2006} whose general form is a superposition of four basis states, $\left\{\ket{HH},\, \ket{HV},\, \ket{VH},\, \ket{VV}\right\}$:
\begin{equation}
    \ket{\Psi}=C_{1}\ket{HH} + C_{2}\ket{HV} + C_{3}\ket{VH}+C_{4}\ket{VV},
\end{equation}
where complex coefficients $C_{1,2,3,4}$ satisfy the normalization condition $|C_{1}|^{2}+|C_{2}|^{2}+|C_{3}|^{2}+|C_{4}|^{2}=1$.
To conduct two-photon polarization tomography, we added a QWP, a HWP and a PBS cube also to Arm 1 of the NPBS, enabling the projection of the photons into any given polarization state.
Subsequently, we measured the coincidence rates of photon pairs for 16 different combinations of polarization projections~\cite{White_PRL_1999} (Table \ref{tab:two_tomo}) and implemented the maximum likelihood state estimation~\cite{Bogdanov_PRA_2006} to extract the density matrix $\hat{\rho}$ of the two-photon state.

We first performed the aforementioned procedures for the single-photon tomography.
As described earlier, we deterministically separated the photons based on their frequencies: for photons pairs generated by MD-qBIC on QOM-A [Figure~\ref{fig:SPDC}(b)], we placed a longpass (LP) filter with the cut-on wavelength at 1550 nm in one of the arms after the NPBS [Fig.~\ref{fig:SPDC}(a)]. 
The LP filter placed in Arm 1 enabled the single-photon tomography of the resonant (signal) photon. Conversely, with the LP filter in Arm 2, the tomography applied to the idler photon.
The other arm served for heralding.
Similarly, for photon pairs generated by the ED-qBIC of QOM-B [Figure~\ref{fig:SPDC}(c)], we placed a 50 nm FWHM bandpass (BP) filter centered at 1475 nm in one of the arms.
In this case, the BP filter placed in Arm 1 enabled the tomography of the idler photon, whereas with the BP filter in Arm 2, the tomography applied to the resonant (signal) photon. 
Note that this method has the drawback of discarding half of the two-photon events.

\begin{figure}[t]
\centering
\includegraphics[width=12cm]{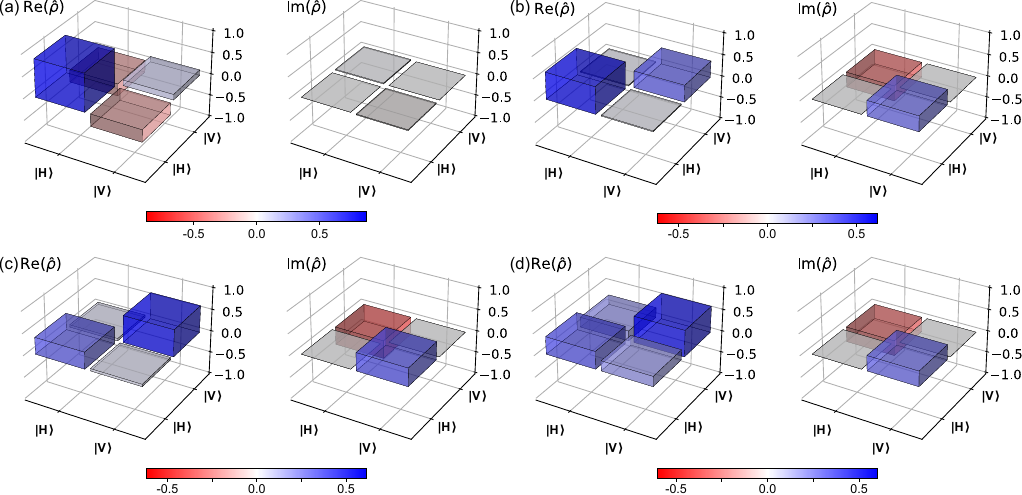}
\caption{\textbf{Single-Photon Polarization Tomography} (a) Real and imaginary parts of the density matrix $\hat{\rho}$ for the signal photon and (b) for the idler photon, respectively, when photon pairs are generated by the MD-qBIC of QOM-A. (c) Real and imaginary parts of the density matrix for the signal photon, and (d) for the idler photon, respectively, when photon pairs are generated by the ED-qBIC of QOM-B.}
\label{fig:SinglePhotonTomo}
\end{figure}

Analysis of the reconstructed density matrices (Fig.~\ref{fig:SinglePhotonTomo}) reveals that the signal photons from MD-qBIC and ED-qBIC are in very distinct polarization states, which are closely matched to the far-field polarization of the qBIC modes [Fig.~\ref{fig:Schematic}(e and d)]. Indeed, photons at MD-qBIC are almost purely horizontally polarized [comp. Fig.~\ref{fig:SinglePhotonTomo}(a) and Fig.~\ref{fig:Schematic}(e)], while photons at ED-qBIC are diagonally polarized [comp. Fig.~\ref{fig:SinglePhotonTomo}(c) and Fig.~\ref{fig:Schematic}(d)].
This indicates that the polarization states of the signal photons can be tailored by modifying the chosen qBIC mode or the symmetry of the meta-atoms.

From each density matrix, we find the purity of the single-photon state, defined as $P$=Tr$\left[\hat{\rho}^{2}\right]$ and corresponding to the degree of polarization in classical optics.
For photon pairs generated by the MD-qBIC of QOM-A, the measured purity (degree of polarization) values for signal and idler photons, respectively, are $0.97\pm0.03 (0.94\pm0.09)$ and $0.80\pm0.05 (0.77\pm0.11)$. 
Similarly, for photon pairs generated by the ED-qBIC of QOM-B, the  purity (degree of polarization) values for the signal and idler photons are $0.92\pm0.05 (0.90\pm0.10)$  and $0.72\pm0.05 (0.72\pm0.10)$, respectively.
We notice that the purity of the signal photons is consistently higher than that of the idler photons.
This difference originates from the involvement of multiple low-$Q$ modes in the generation of the idler photons.
Indeed, as explained above [see Fig.~\ref{fig:Schematic}(c)], the QOMs support in-plane ED and MD bright modes, too.
These modes span the spectral range where the idler photons are generated.
The signal photons, which originate from a high-$Q$ qBIC resonance, have a very long lifetime and a defined polarization state $\left(\ket{P_{s}}\right)$ given by the mode.
On the other hand, idler photons are generated at wavelengths that can populate either of the orthogonally polarized $(\ket{P_{1}}$ and $\ket{P_{2}})$ bright modes [Fig.~\ref{fig:Schematic}(f, g)], and therefore, are in a superposition of these two orthogonal low-$Q$ modes.
Consequently, during the lifetime of the signal photon, the idler photon is in a mixed state, with the two-photon state being $\ket{P_{s}}\bigotimes\left(\alpha\ket{P_{1}}+\beta\ket{P_{2}}\right)$.
Here, $\alpha$ and $\beta$ are the probability amplitudes of populating $\ket{P_{1}}$ and $\ket{P_{2}}$, respectively.
This behavior is similar to the one of an SPDC source where one of the two photons is confined in a cavity~\cite{Burlakov_LaserPhys_1996}.
Our finding suggests that to have both photons within the pair with high purity, the QOMs should support a single mode for each generated photon.
An alternative way is to simply utilize  photon pairs from a degenerate SPDC process, in which two photons will be emitted with the same polarizations.

\begin{figure}[t]
\centering
\includegraphics[width=12cm]{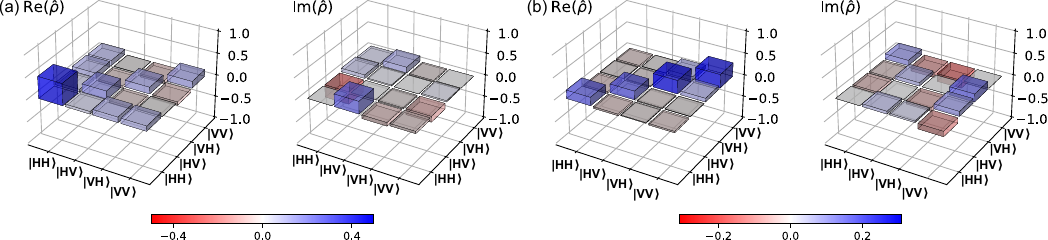}
\caption{\textbf{Two-Photon Polarization Tomography} (a) Real and imaginary part of the density matrix for photon pairs generated by the MD-qBIC on QOM-A and (b) by the ED-qBIC on QOM-B, respectively.}
\label{fig:TwoPhotonTomo}
\end{figure}

Further, we performed two-photon polarization tomography, where we installed the QWP, HWP and PBS in both arms of the setup.
For the characterization of the reconstructed density matrices, we computed their purity and the entanglement of formation~\cite{Bennett_PRA_1996, Wootters_PRL_1998}, which describes the degree of entanglement in a two-photon polarization state.
The two-photon polarization state generated by the MD-qBIC [Fig.~\ref{fig:TwoPhotonTomo}(a-b)] exhibited a purity of 0.71$\pm$0.03 and an entanglement of formation of 0.14.
Meanwhile, the two-photon polarization state generated by the ED-qBIC [Fig.~\ref{fig:TwoPhotonTomo}(a-b)] had a purity of 0.47$\pm$0.06 and an entanglement of formation of 0.01.
In both cases, the two-photon polarization states were in a mixed state and nearly separable, which is a direct consequence of the generated photons having a well-defined polarization according to the the way in which we break the symmetry of the meta-atom.
The results of all polarization tomography measurements are summarized in Table \ref{tab:pol_tomo}.

\begin{table}[h!]
\centering
\begin{tabular}{cccccccc}
\hline
 \textbf{Polarization} & \textbf{Measured qBIC} & \textbf{Purity} & \textbf{Degree of} & \textbf{Entanglement of } \\ 
  \textbf{Tomography Type} & \textbf{and selected photon} & &  \textbf{Polarization} & \textbf{ Formation}   \\ 
\hline
Single-Photon & MD-qBIC signal photon& 0.97$\pm$0.03 & 0.94$\pm$0.09 & N/A \\\
Single-Photon & MD-qBIC idler photon & 0.80$\pm$0.05 & 0.77$\pm$0.11 & N/A \\\
Single-Photon & ED-qBIC signal photon & 0.92$\pm$0.05 & 0.90$\pm$0.10 & N/A \\\
Single-Photon & ED-qBIC idler photon & 0.72$\pm$0.05 & 0.72$\pm$0.10 & N/A \\\
Two-Photon & MD-qBIC & 0.71$\pm$0.03  & N/A & 0.14 \\\
Two-Photon &  ED-qBIC & 0.47$\pm$0.06  & N/A & 0.01 \\\
\end{tabular}
\caption{Summary of the polarization tomography measurements.} 
\label{tab:pol_tomo}
\end{table}

\begin{figure}[h!]
\centering
\includegraphics[width=12cm]{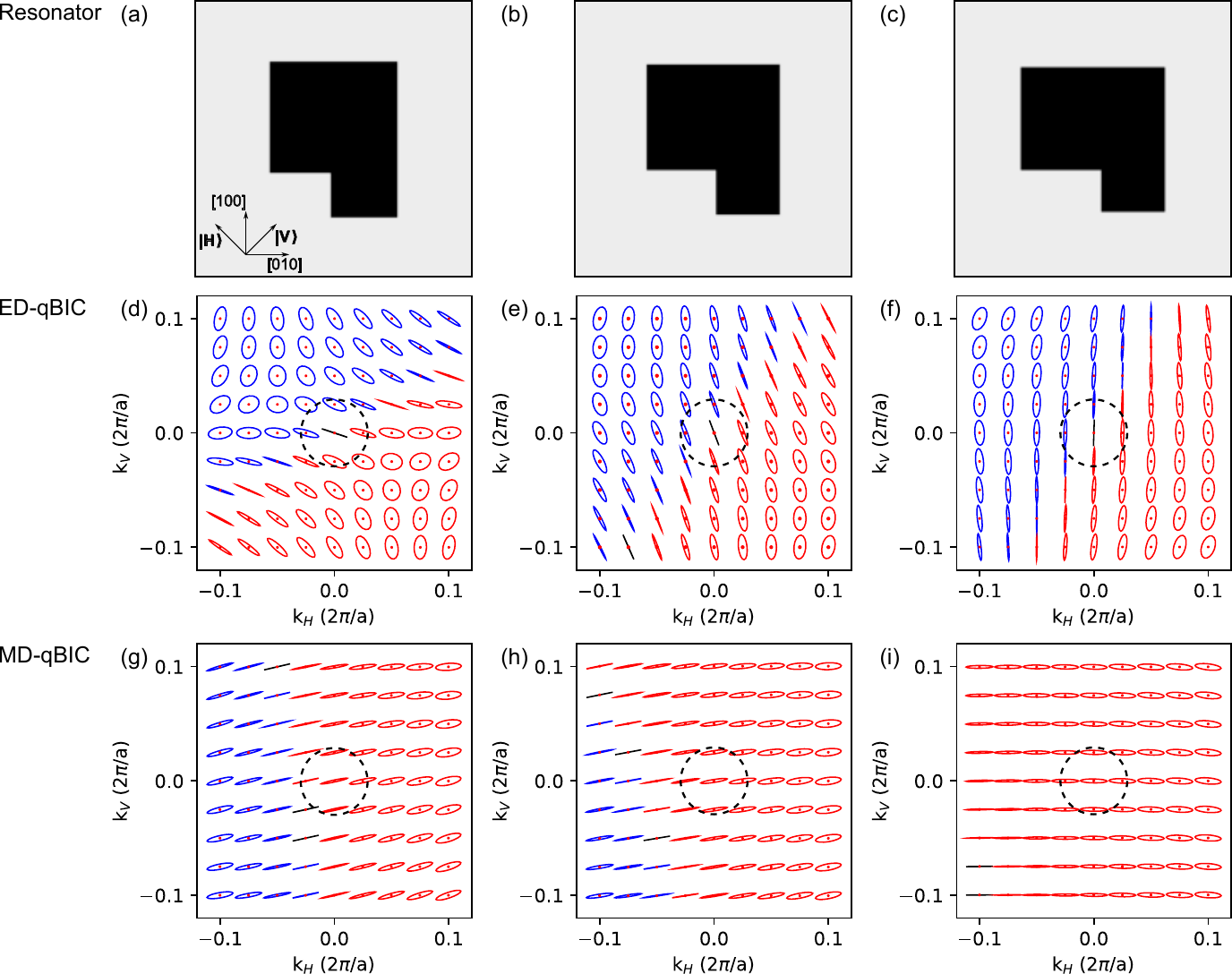}
\caption{(a-c) Shapes of the meta-atoms gradually varied in steps from a meta-atom where a rectangular protrusion is added to a square, as shown in Fig. \ref{fig:Schematic}(a), to a meta-atom where a rectangular notch is removed from a square, as shown in (c). The total area of the meta-atom and the periodicity of the square lattice are kept constant in the simulation. (d-f) The far-field polarization states of the metasurface for the bands on which ED-qBIC and (g-i) MD-qBIC lie at the $\Gamma$ point, where individual meta-atoms of the metasurface correspond to (a-c), respectively. The blue (red) ellipses represent right-handed (left-handed) polarization states. Dashed circles indicate the range in $k$-space covered by the effective collection NA of $\sim 0.03$.
}
\label{fig:FarFieldBIC}
\end{figure}

Our approach of generating the SPDC photon pairs via metasurfaces allows for complete freedom in meta-atom design, which eliminates any restrictions in utilizing symmetries or lack of symmetries to fully engineer the polarizations of the resonant modes.
In Fig.~\ref{fig:FarFieldBIC}, we present a straightforward modification of our broken-symmetry meta-atoms, which show great potential for the subsequent polarization engineering.
Here, we start from the meta-atom geometry studied in Fig.~\ref{fig:Schematic}(a) and gradually modify the aspect ratio of the side lengths, transforming it into the final shape displayed in Fig.~\ref{fig:FarFieldBIC}(c), which essentially is a square with a rectangular notch.
Employing a GME simulation, we calculate the far-field polarizations of both ED-qBIC and MD-qBIC modes for these progressively asymmetric shapes of the meta-atom (results in Fig.~\ref{fig:FarFieldBIC}, second and third rows, respectively).
We observe a dramatic rotation in the far-field polarization of the ED-qBIC modes compared to the MD-qBICs, which maintain a more stable polarization predominantly aligned along the H-polarization axis. 
This tunability highlights the potential for engineering specific polarization states by playing with the degree of asymmetry in the meta-atoms.
In addition to manipulating the symmetry of the meta-atom, the calculated far-field polarization patterns show that we can further control the polarization characteristics of these photons by manipulating the NA of the collection, enabling the generation of mixed or pure single photons.
For example, we can control the polarization state of resonantly emitted photons: a very low collection NA, as used in our experiment, yields pure polarization states, while increasing the NA would result in mixed polarization states.

In summary, we have demonstrated the first comprehensive polarization tomography of entangled photon pairs generated via nondegenerate SPDC in nonlinear metasurfaces.
We identified that the two-photon polarization states exhibit characteristics of mixed and nearly separable states.
Furthermore, we observed that the single-photon polarization state of the signal photons closely resembles the far-field polarization of the qBICs, and exhibits higher purity than the idler photons' state. This behavior is in stark contrast to the one of photon pairs generated in bulk crystals: there, either both photons are in mixed states and the pair is entangled, or both photons are pure and the pair is factorable. A potential application of such metasurfaces is heralded generation of single photons with controllable polarization.
In this regard, an intriguing direction for future investigation involves the utilization of GaAs with different crystallographic orientations, such as (110)-GaAs, which not only holds promise for enhancing the SPDC generation rate, but also provides a control knob in addition to the far-field polarization of the qBICs to engineer the quantum state~\cite{Sultanov_OptLett_2022}.
Previously, complex schemes relying on multiple sets of crystals or temporal manipulation of pump beam decoherence were proposed to produce arbitrary two-photon polarization mixed states~\cite{Wei_PRA_2005}.
In contrast, our work provides a significantly simplified scheme for achieving an arbitrary two-photon state by pre-determining them by controlling the qBIC mode profiles.

\section*{Data availability}
The data that support the findings of this study are available from the corresponding author on reasonable request.

\section*{Acknowledgements}
This work was supported by the U.S. Department of Energy (DOE), Office of Basic Energy Sciences, Division of Materials Sciences and Engineering and performed, in part, at the Center for Integrated Nanotechnologies, an Office of Science User Facility operated for the U.S. DOE Office of Science.
Sandia National Laboratories is a multi-mission laboratory managed and operated by National Technology and Engineering Solutions of Sandia, LLC., a wholly owned subsidiary of Honeywell International, Inc., for the U.S. DOE’s National Nuclear Security Administration under contract DE-NA0003525.
This paper describes objective technical results and analysis. Any subjective views or opinions that might be expressed in the paper do not necessarily represent the views of the U.S. DOE or the United States Government.
V.S. and M.V.C are part of the Max Planck School of Photonics supported by BMBF, Max Planck Society and Fraunhofer Society.
This work was funded by the Deutsche Forschungsgemeinschaft (DFG, German Research Foundation) – Project-ID 429529648 – TRR 306 QuCoLiMa (“Quantum Cooperativity of Light and Matter’’).

\section*{Competing interests}
The authors declare no competing interests.
\clearpage

\pagebreak
\widetext
\begin{center}
\textbf{\large Supplemental Materials: Quantum Pair Generation in Nonlinear Metasurfaces with Mixed and Pure Photon Polarizations}
\end{center}
\setcounter{equation}{0}
\setcounter{figure}{0}
\setcounter{table}{0}
\setcounter{page}{1}
\makeatletter
\renewcommand{\theequation}{S\arabic{equation}}
\renewcommand{\thefigure}{S\arabic{figure}}
\renewcommand{\thetable}{S\arabic{figure}}

\section*{I. Fiber-assisted SPDC spectroscopy}
\begin{figure}[htbp]
\centering
\includegraphics[width=8.4cm]{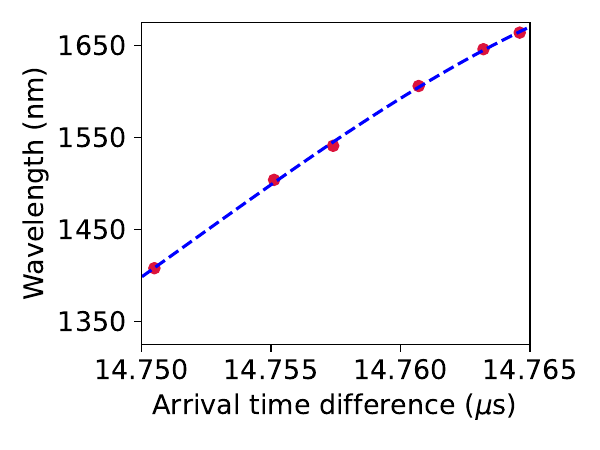}
\caption{The conversion of the arrival time differences of the photons at the two detection arms to the wavelength of one of them (the other wavelength being defined by the energy conservation) calibrated using different spectral filters. The dashed line represents the fit with a polynomial function.}
\label{fig:sup_spectroscopy}
\end{figure}

Given the low rate at which SPDC is generated, conducting a direct spectroscopy is challenging.
To address this limitation, we adopt a method known as fiber-assisted SPDC spectroscopy\cite{Valencia_PRL_2002, Santiago-Cruz_Science_2022, Sultanov_Nature_2024}, which relies on coincidence measurements.
In addition to the experimental setup for the coincidence measurements, we inserted a 3km spool of single-mode fiber (Corning SMF-28) in one of the detection arms (arm 1) for measuring the  spectrum of SPDC photons.
Photons traveling through arm 1 experience wavelength-dependent time delays due to the group velocity dispersion of the fiber, resulting in the broadening of the coincidence count histogram in time.
Using a calibration curve derived from various spectral filters (Fig. \ref{fig:sup_spectroscopy}), we convert the differences in the arrival times of the two photons into these wavelengths of the photons.
The detectors' timing jitter sets the spectral resolution, which results in being below 3 nm.
Furthermore, the combinations of spectral filters used for each SPDC spectroscopy measurement are shown in Table \ref{tab:spectroscopy}.

\begin{table}[htbp]
\centering
\begin{tabular}{cccccccc}
\hline
 \textbf{Measured SPDC photons} & \textbf{Before NPBS} & \textbf{Arm 1} & \textbf{Arm 2} \\ 
\hline
MD-qBIC signal and idler photons & LP1250 + 2$\times$LP1400 + LP1500 & None & None \\
ED-qBIC signal photon & LP1250 + 2$\times$LP1400 & BP1475 & None \\
ED-qBIC idler photon & LP1250 + 2$\times$LP1400 & None & BP1475 \\
\end{tabular}
\caption{Spectral Filters used for each SPDC spectroscopy measurement. LP1250, LP1400 and LP1500 are long-pass filters for 1250nm, 1400nm, and 1500nm, respectively, and BP1475 is a bandpass filter at 1475nm (50 nm FWHM).} 
\label{tab:spectroscopy}
\end{table}

\begin{figure} [htbp]
\includegraphics[width=14cm]{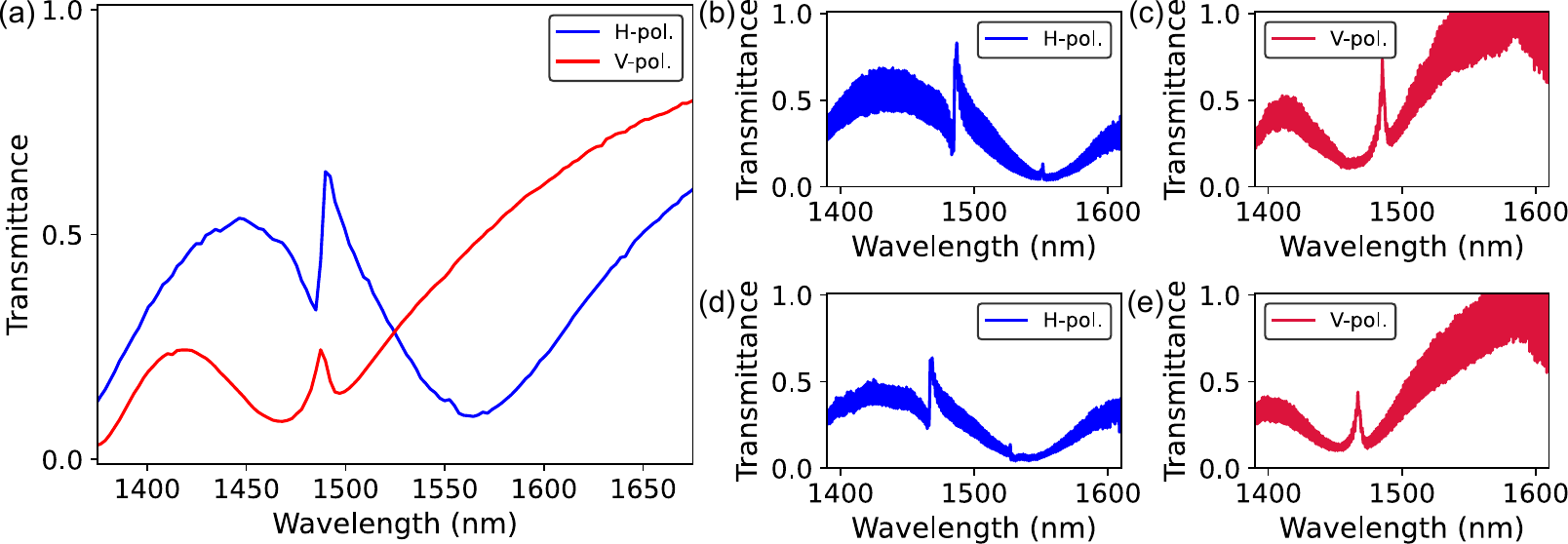}
\caption{(a) Measured white light transmittance of QOM-B for the incident polarization along the H-polarization axis (blue) and V-polarization axis (red) with an Ocean Optics NIRQuest IR spectrometer.
(b) Measured white light transmittance of QOM-B measured with higher resolution Czerny–Turner spectrograph (Acton Pro, Spectra-Pro 2500i) for the incident polarization along the H-polarization axis and (c) V-polarization axis, respectively, and (d) of QOM-A for the incident polarization along the H-polarization axis and (e) V-polarization axis, respectively.}
\label{fig:lin}
\end{figure}

\section*{II. Polarization projections for the density matrix reconstruction.}
\label{supp:polprojection}
The specific details of the standard protocol for the polarization tomography can be found in D.F.V. James et al. \cite{James_PRA_2001} and V. Sultanov et al. \cite{Sultanov_Nature_2024}.
The polarization projections of the single- and two-photon polarization tomography are presented in Tables \ref{tab:single_tomo} and \ref{tab:two_tomo}, respectively.
In addition, Stokes parameters of the single-photon polarization states can be calculated from the same measurements. Each of the Stokes parameters can be directly deduced from the projection measurements as\cite{Chekhova_Banzer_2021}: $S_{0}=P_{H}+P_{V}$, $S_{1}=P_{H}-P_{V}$, $S_{2}=P_{D}-P_{A}$, and $S_{3}=P_{L}-P_{R}$, where $P_{\ket{i}}$ is the probability to measure the state $\ket{i}$.
Furthermore, we can compute the degree of polarization from the Stokes parameters as $P\equiv\frac{\sqrt{S_{1}^{2}+S_{2}^{2}+S_{3}^{2}}}{S_{0}}$, which ranges from 0 (maximally mixed states) to 1 (pure states), with partially mixed states in between.
This contrasts with the purity of the density matrix of an $n$-photon polarization state, which ranges from $1/2^{n}$ (maximally mixed states) to 1 (pure states).

\begin{table}[htbp]
\centering
\begin{tabular}{cccccccc}
\hline
 \textbf{Polarization state} & \textbf{HWP}  & \textbf{QWP} \\  
\hline
H & 0$^{\circ}$ & 0$^{\circ}$ \\\
V & 45$^{\circ}$ & 0$^{\circ}$ \\\
D & 22.5$^{\circ}$  & 45$^{\circ}$ \\\
A & -22.5$^{\circ}$  & 45$^{\circ}$ \\\
R & 0$^{\circ}$  & -45$^{\circ}$ \\\
L & 0$^{\circ}$  & 45$^{\circ}$ \\\
\end{tabular}
\caption{6 projection measurements required for the full single-photon polarization tomography.} 
\label{tab:single_tomo}
\end{table}
\begin{table}
\centering
\begin{tabular}{cccccccc}
\hline
 \textbf{Polarization state} & \textbf{HWP$_{1}$}  & \textbf{QWP$_{1}$} & \textbf{HWP$_{2}$} & \textbf{QHP$_{2}$} \\  
\hline
H-H & 0$^{\circ}$ & 0$^{\circ}$ & 0$^{\circ}$  & 0$^{\circ}$ \\\
H-V & 0$^{\circ}$ & 0$^{\circ}$ & 45$^{\circ}$  & 0$^{\circ}$ \\\
V-H & 45$^{\circ}$ & 0$^{\circ}$ & 0$^{\circ}$  & 0$^{\circ}$ \\\
V-V & 45$^{\circ}$ & 0$^{\circ}$ & 45$^{\circ}$  & 0$^{\circ}$ \\\
H-D & 0$^{\circ}$ & 0$^{\circ}$ & 22.5$^{\circ}$  & 45$^{\circ}$ \\\
H-R & 0$^{\circ}$ & 0$^{\circ}$ & 0$^{\circ}$  & -45$^{\circ}$ \\\
V-A & 45$^{\circ}$ & 0$^{\circ}$ & -22.5$^{\circ}$  & 45$^{\circ}$ \\\
V-L & 45$^{\circ}$ & 0$^{\circ}$ & 0$^{\circ}$  & 45$^{\circ}$ \\\
D-H & 22.5$^{\circ}$  & 45$^{\circ}$ & 0$^{\circ}$ & 0$^{\circ}$ \\\
R-H & 0$^{\circ}$  & -45$^{\circ}$ & 0$^{\circ}$    & 0$^{\circ}$ \\\
A-V  & -22.5$^{\circ}$  & 45$^{\circ}$ & 45$^{\circ}$ & 0$^{\circ}$\\\
L-V  & 0$^{\circ}$  & 45$^{\circ}$ & 45$^{\circ}$ & 0$^{\circ}$\\\
D-D & 22.5$^{\circ}$  & 45$^{\circ}$ & 22.5$^{\circ}$  & 45$^{\circ}$ \\\
R-R & 0$^{\circ}$  & -45$^{\circ}$ & 0$^{\circ}$  & -45$^{\circ}$ \\\
D-R & 22.5$^{\circ}$  & 45$^{\circ}$ & 0$^{\circ}$  & -45$^{\circ}$ \\\
R-D & 0$^{\circ}$  & -45$^{\circ}$ & 22.5$^{\circ}$  & 45$^{\circ}$ \\\
\end{tabular}
\caption{16 projection measurements required for the full two-photon polarization tomography. Subscripts 1 and 2 indicate each pair of waveplates placed in front of the detectors on each arm.} 
\label{tab:two_tomo}
\end{table}
\bibliography{reference}

\end{document}